\documentclass[10pt]{article}

\usepackage{amsmath}
\usepackage{amssymb, chemarr}

\usepackage{graphicx}

\usepackage{cite}

\usepackage{color} 

\usepackage[labelfont=bf,labelsep=period,justification=raggedright]{caption}

\makeatletter
\renewcommand{\@biblabel}[1]{\quad#1.}
\makeatother

\date{}

\begin{document}

\begin{flushleft}
{\Large
\textbf{Measuring the Degree of Modularity from Gene Expression Noise in Gene Regulatory Circuits}
}
\\
Kyung Hyuk Kim$^{1, \ast}$, 
Herbert M. Sauro$^{1}$, 
\\
\bf{1} Department of Bioengineering, University of Washington, Seattle,
WA 98195, USA
\\
$\ast$ E-mail: kkim@u.washington.edu
\end{flushleft}

\section*{Abstract}
Gene regulatory circuits show significant stochastic fluctuations in their circuit signals due to the low copy number of transcription factors.  When a gene circuit component is connected to an existing circuit, the dynamic properties of the existing circuit can be affected by the connected component.  In this paper, we investigate modularity in the dynamics of the gene circuit based on stochastic fluctuations in the circuit signals.  We show that the noise in the output signal of the existing circuit can be affected significantly when the output is connected to  the input of another circuit component.  More specifically, the output signal noise can show significantly longer correlations when the two components are connected.  This equivalently means that the noise power spectral density becomes narrower.   We define the relative change in the correlation time or the spectrum bandwidth by stochastic retroactivity, which is shown to be directly related to the retroactivity defined in the deterministic framework by del Vecchio et al. This provides an insight on how to measure retroactivity, by investigating stochastic fluctuations in gene expression levels, more specifically,  by obtaining an autocorrelation function of the fluctuations.   We also provide an interesting aspect of the frequency response of the circuit.  We show that depending on the magnitude of operating frequencies, different kinds of signals need to be preferably chosen for circuit description in a modular fashion: at low enough frequency, expression level of transcription factor that are not bound to their specific promoter region needs to be chosen, and at  high enough frequency, that of the total transcription factor, both bound and unbound, does.


\section*{Introduction}

In biology, the word modularity has multiple meanings depending on context. It can for example represent a set of co-expressed genes~\cite{stuart2003gene}, a topological unit in a network~\cite{girvan2002community} or a self-contained component such as an organ. In this paper we will define a module as a self-contained functional unit whose intrinsic properties are independent of the surrounding milieu. This definition is essentially the same definition used in engineering. For example, the intrinsic properties of a TTL NAND gate~\cite{TTLCookbook} is unaffected (within certain design constraints) when connected to other TTL logic gates. It is the property that makes the engineering of complex electronic circuits possible. It allows engineers to design, predict and fabricate with a high degree of reliability. The question whether such self-contained and functionally independent modules exist at the biological cellular network level is still an ongoing research problem \cite{Alon2007}. In this article we will be concerned with the design of modular synthetic components. Given the recent rise in interest in synthetic biology \cite{Purnick2009, Keasling2008, Voigt2006, Sprinzak2005, Endy2005}, the ability to design and build novel cellular systems in a modular fashion is a highly desirable goal.

\begin{figure}[h]
  \centering
  \includegraphics[width=3.3in]{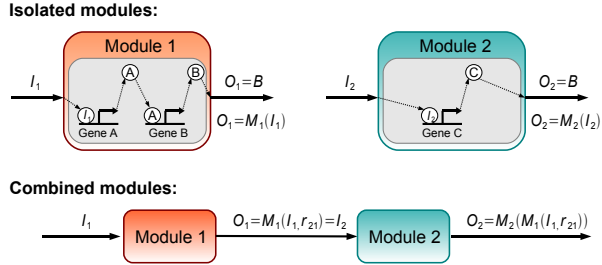}
  \caption{Gene circuit modules: Input ($I$)  and output ($O$) signals are TF concentrations.  The I/O relationships of the modules, $M_1$ and $M_2$, are given by the solutions of  ordinary differential equations, describing the temporal changes in the concentration levels of the TFs, in the deterministic case.  Module 2 affects the dynamics of module 1 due to retroactivity on module 1 by module 2, $r_{21}$.} \label{fig:module}
\end{figure}

In the most abstract sense we can define a module as follows. Given a functional unit $M$ with input $I$ and output $O$, we can define a relation between the input and output as $O = M(I)$. Given two functional units, $M_1$ and $M_2$, where the output of $M_1$ serves as the input to $M_2$, then $M_1$ and $M_2$ are defined as modules if the relation, $O_2 = M_2 (M_1 (I_1))$ is true (Fig.~\ref{fig:module}). This simply means that in connecting $M_1$ and $M_2$ together, $M_2$ has no effect on the functional characteristics of $M_1$ and vice versa.   

Del Vecchio et al. \cite{Vecchio2008} introduced a measure called retroactivity that allowed one to determine the influence that a downstream module had on the characteristics of an upstream module. The higher the retroactivity the greater the influence from the downstream module and the less modular the functional upstream units. More specifically, they considered a gene circuit module where the input output signals were the concentrations of transcription factors, wiring the module to another (see Fig.~\ref{fig:module}).  It was shown that a module could respond to perturbations much more slowly when the output was wired to the input of the next module~\cite{Vecchio2008}.  The wiring involves binding-unbinding processes between TFs and their specific promoter regions.  The lifetime of the bound TFs was assumed to be much larger than that of the unbound TFs.  If the processes, initially at equilibrium, are perturbed in such a way that the copy number of the free TF is abruptly  increased, a portion of the increased amount will bind to the promoter region if any regions are not occupied. This results in a quasi-equilibrium in the binding-unbinding process \cite{Kepler2001, Rao2003}.  The promoters act as a reservoir holding the TFs: Whenever there is any change in the number of the free TFs, the reservoir quickly buffers the change in the number of TFs \cite{Buchler2009}.  The slow-down in the dynamic response in the upstream circuit component was quantified by a retroactivity measure \cite{Vecchio2008}.   Del Vecchio et al. proposed a mechanism for reducing the retroactivity to achieve modular analysis of the circuit dynamics.

In this manuscript, we focus on an operational method for measuring the retroactivity, by exploiting gene expression noise.    The gene circuits have been known to show significant stochastic fluctuations in their gene expression levels \cite{Arkin1998, Elowitz2002, Ozbudak2002,Elf2007,Rosenfeld2005, Austin2006,  Weinberger2008, Dunlop2008} (for review articles, \cite{Rao2002, Raser2004, Kaern2005, Shahrezaei2008b}).   Stochastic noise often involves useful information that is not available by observing the mean values \cite{Munsky2009}.  The noise in the gene expression can be considered as an outcome of continuous perturbations applied to the gene circuits.  Thus, the system's dynamic properties   can be obtained from the noise.   In this manuscript,  we investigate the measurable changes in the noise statistics due to connections between gene circuit components and propose a novel practical method  to measure retroactivity \emph{in vivo} to quantify the level of modularity of the dynamics of the circuit components.  

We show that  the output signal noise can exhibit a significantly longer correlation in the output signal when the signal is used as an input of a downstream circuit component.  The change in the temporal correlation will be quantified by an autocorrelation function \cite{Anishchenko} (for gene expression studies, \cite{Rosenfeld2005, Austin2006, Weinberger2008, Dunlop2008}), which indicates the correlation of the signal ($X$) with itself for a given time-lag ($\tau$):
\begin{equation}
G_X(\tau) = \big\langle (X_{t+\tau} -\langle X \rangle)(X_t- \langle  X \rangle) \big\rangle,
\label{eqn:autocorr-def}
\end{equation}
where $X_t$ is the signal amplitude at time $t$ and is presumed at a stationary state fluctuating with respect to a constant mean.   The angle bracket denotes an average over a time series.   The observed longer correlation in the signal means that the autocorrelation will decrease more slowly with the time-lag, which  equivalently means that the signal noise has a narrower bandwidth in the frequency spectrum.  We can quantify such changes in the autocorrelation or the frequency spectrum by introducing a stochastic retroactivity.  We show that this stochastic retroactivity (s-retroactivity) is directly related to the retroactivity investigated in the deterministic framework (d-retroactivity) \cite{Vecchio2008}.   

We also investigate the change in the frequency response of the gene circuit due to the circuit connection by observing the autocorrelation functions, more specifically power spectral densities -- the Fourier transforms of the autocorrelation functions.  If circuit signals can be chosen such that they are less affected by the connection, the chosen signals would be preferred by modelers since the circuit dynamics can be analyzed in a modular way.  We show that the proper choice is dependent of the magnitude of the frequency where the circuits are operated.  Our analysis on the power spectral densities  show that  the concentrations of the free TFs are not affected by the wiring at low frequencies (of the order of a day) circuit operation, while those of the total TFs, including free and bound,  at high frequency (of the order of an hour) operation.  This implies that depending on the magnitude of the operating frequency, the signal output can be chosen either the free or total TFs to describe its dynamics in a modular fashion.  

Autocorrelation functions have been experimentally measured to understand the noise power spectrum in gene expression levels of a HIV transcriptional circuit \cite{Weinberger2008} and TetR negative feedback circuit \cite{Austin2006}. Autocorrelation has also been used to investigate the properties of intrinsic and extrinsic noise \cite{Rosenfeld2005} and to analyze regulatory interactions in a CRP-GalS-GalE feed-forward circuit \cite{Dunlop2008}. Here we offer a new application of autocorrelation functions in relation to the analysis of modularity.

\section*{Results}
\subsection*{Retroactivity: Relative increase in the correlation time of output signal}
\begin{figure}[h!]
  \centering
  \includegraphics[width=3.in]{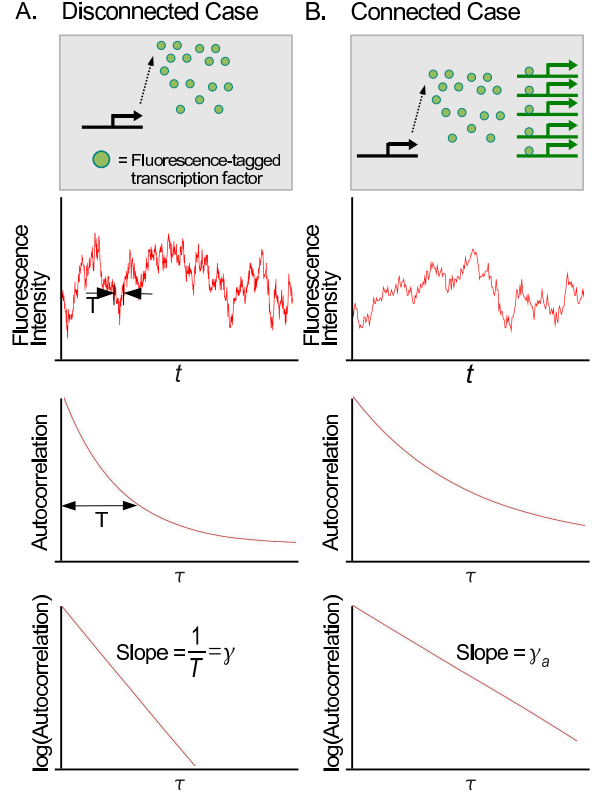}
  \caption{Stochastic fluctuations in a fluorescence signal (denoted by $Y(t)$) and its autocorrelation function $G_Y(\tau)$. The signal $Y(t)$ fluctuates stochastically with respect to its constant mean level.  An autocorrelation, indicating the correlation of a signal with itself at a given time difference, shows that  the correlation typically decreases as the time difference $\tau$ increases.  If the autocorrelation function decreases exponentially with  $\tau$, i.e., $G_Y (\tau) = G_Y(0) e^{-\frac{\tau}{T}}$, the correlation time constant is defined by $T$.  This time constant is roughly the same as the time taken for a perturbation in the signal to be relaxed to the the mean level.  When the signal is used for the input of another circuit component (B.~Connected Case), the signal $Y$ shows longer correlations in time. The correlation time increases and the apparent degradation rate constant decreases.} 
\label{fig:autocorr}
\end{figure}

Let us define stochastic retroactivity based on the change in the autocorrelation function of the output signal noise of a module due to the connection of the output to the input of another module.  Consider the translation and degradation process of a TF protein ($X$). We presume this process serves as a simple module, or a portion of a module processing the module output signal ($X$).  We will compare two cases: the module output is wired and not wired (see Fig.~\ref{fig:autocorr}).   We first consider the case when not wired: No wiring means that the TF cannot bind to a TF-specific promoter region, e.g., perhaps the TF specific binding regions of promoters are absent.  In such cases, all of the TFs are in an unbound state.  If the TF has a fluorescence tag, the fluorescence intensity will reflect the copy number of the unbound TFs.  We can model the translation-degradation process of the TF as:  
\begin{equation}
\xrightarrow{~~\alpha~~} X \xrightarrow{\gamma X}{} \varnothing,
\label{eqn:isolated}
\end{equation}
where $\alpha$ is a translation rate, and $\gamma X$ a degradation rate.  We neglect a dilution effect (due to cell growth) and other extrinsic noise sources, which however will be taken into account later in this section. 

Consider the above reaction system in the deterministic framework.  When the number of $X$ is perturbed to increase from the stationary value, the characteristic time to reach the stationary state is $1/\gamma$ \cite{note-retroact}.  In the stochastic description, stochastic fluctuations in $X$, deviated from the stationary state mean value, will spend a time $1/\gamma$ typically in reaching the mean value (see Fig.~\ref{fig:autocorr}).  This means that the autocorrelation function decreases with a time constant $1/\gamma$, more specifically, the autocorrelation can be shown to decrease exponentially: $G_X(\tau) = G_X(0) e^{-\gamma \tau}$ \cite{Anishchenko}.  After applying logarithms on both hand sides, we obtain $\log G_X(\tau) = \log G_X(0) - \gamma \tau$.  The semi-log plot of $G_X(\tau)$ becomes a linear line with its slope $-\gamma$, which is  the degradation rate constant up to the negative sign. Thus, this implies that the deterministic retroactivity (d-retroactivity), quantifying the slow-down in response due to wiring  \cite{Vecchio2008}, can be measured from the slope change.

We will next wire the output to another module (see Fig.~\ref{fig:autocorr}).  Wiring means allowing $X$ to bind to  the promoter region of a downstream module, either by inserting specific promoters for $X$ or by adding inducer if the promoter was already present but its activity was suppressed.   
The system with the connection can be modeled as:
\begin{equation}
\xrightarrow{~~\alpha~~}X \xrightarrow{\gamma X} \O,~~\mbox{and}~~X+P_f\xrightleftharpoons[k_{off}P_b]{k_{on}XP_f}P_b,
\label{eqn:rate}
\end{equation}
where $P_f$ and $P_b$ denote free and bound promoters, respectively. The binding-unbinding process can be considered much faster in its relaxation to the stationary state than the translation-degradation process ($k_{on}X + k_{off} \gg \gamma$; cf. \cite{Kepler2001, Simpson2004}) and is assumed to be in quasi-steady states.  

In this subsection, we will focus more on the total number of the TFs rather than the number of the free TFs ($X$).  The reason is that the total number can be experimentally observable if the fluorescence does not change when the TFs are bound, and we will denote the total number by 
\[
Y\equiv X+ P_b.
\]
If the TFs have fluorescence tags, they will bind to the promoter region and the fluorescence intensity will reflect the sum of the copy numbers of both bound and unbound TFs.   The second reason that we focus more on the sum $Y$ is that it directly reflects the dynamics of  the time scale of our interest (of the order of cell-doubling time or less).  In this time scale, the fast binding-unbinding reactions occur many times, and the number fluctuations in the free TFs are rapid, and these fluctuations can be considered averaged out in this time scale.  Thus it is natural to consider a variable that does not fluctuate due to the binding-unbinding process.  The total number $Y$ satisfies this property since the total number of the TFs does not change while a TF binds to or unbinds from its specific promoter region if a TF is not translated and does not degrade yet.   Thus, the variable $Y$ can be considered a pure slow mode \cite{Rao2003, Vecchio2008}.  

The slow mode $Y$ can fluctuate only when TFs are translated and degrade. This indicates that the  translation-degradation process \eqref{eqn:rate} can be equivalently described (whether the binding-unbinding process is fast or not) by
\begin{equation}
\xrightarrow{~~\alpha~~} Y \xrightarrow{\gamma X},~~\mbox{and}~~X+P_f\xrightleftharpoons[k_{off}P_b]{k_{on}XP_f}P_b.
\label{eqn:rate-y}
\end{equation}
This process can be simplified further by using the fact that the binding-unbinding events occur much more frequently than that of the translation-degradation process.   We assume that the fast fluctuations in the free TF number are averaged out in the time scale of the translation-degradation process.  We can consider the degradation rate of $Y$ to be approximately given as   $\gamma X \simeq \gamma (Y-\langle P_b \rangle_Y)$, where $\langle P_b \rangle_Y$ denotes the average copy number of the bound TFs for a given value of $Y$, which can be estimated from the probability distribution function for $P_b$ obtained under the  assumption of equilibrium in the binding-unbinding process (see the  Appendix).  Thus, the above process for $Y$ can be approximated to
\begin{equation}
\xrightarrow{~~\alpha~~} Y \xrightarrow{\gamma (Y-\langle P_b \rangle_Y)}. 
\label{eqn:rate-y1}
\end{equation}

\begin{figure}[h]
  \centering
  \includegraphics[width=3.3in]{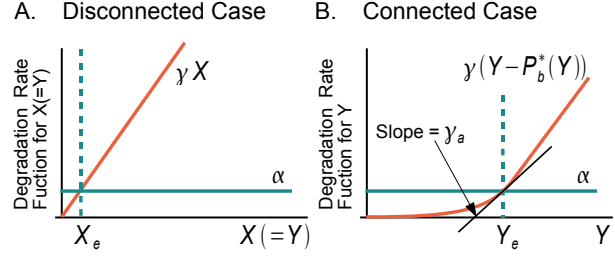}
  \caption{Rate functions for the translation and degradation processes of the total transcription factor $Y$.  The processes are described by \eqref{eqn:isolated} and \eqref{eqn:rate-y1}.  $X_e$ and $Y_e$ are concentrations of the transcription factor at equilibrium.  In the connected case, the degradation rate becomes approximately $\gamma Y$ for sufficiently large $Y$ and the apparent degradation rate constant $\gamma_a$ becomes always smaller than $\gamma$ \cite{Vecchio2008}.} 
\label{fig:degradation}
\end{figure}

Before we discuss how to define retroactivity in the stochastic regime, we will review the deterministic retroactivity (d-retroactivity)~\cite{Vecchio2008}.  Consider the process \eqref{eqn:rate-y1} in the deterministic framework.  The degradation rate of $Y$ is expressed as $\gamma (Y-P_b^*)$, where $P_b^*$ denotes the number of the bound TFs for a given value of $Y$,  estimated under the equilibrium assumption in the binding-unbinding process~\cite{Vecchio2008}.  This degradation rate function was shown to be highly nonlinear in $Y$  \cite{Vecchio2008, Buchler2009}.  We have plotted the graph for the rate function vs. $Y$ in Fig.~\ref{fig:degradation}.   The slope of the graph at $Y_e$, with $Y_e$ the equilibrium value of $Y$, indicates the speed of the relaxation when $Y$ is perturbed from $Y_e$ and thus the slope is the apparent degradation rate constant ($\gamma_a$).   The inverse of the degradation rate constant indicates how long the relaxation lasts.   Since the degradation rate function is highly nonlinear,  the derivative changes significantly for different values of $Y$.  In 
\cite{Vecchio2008}, it was shown that the derivative was always smaller than that of the disconnected case \cite{Vecchio2008}.   Thus, the relaxation time ($1/\gamma_a$) of $Y$ increases significantly compared with the case when disconnected.    The d-retroactivity is defined as the relative change of the apparent  degradation rate constant of $Y$ \cite{Vecchio2008}: 
\begin{equation}
R_d  \equiv  \frac{\gamma - \gamma_a}{\gamma} =  \frac{1}{1+\frac{\left(  1+\frac{X}{K_d} \right)^2}{\frac{P_t}{K_d}}}.
\label{eqn:Rd}
\end{equation}

\begin{figure}[t]
  \centering
  \includegraphics[width=3.3in]{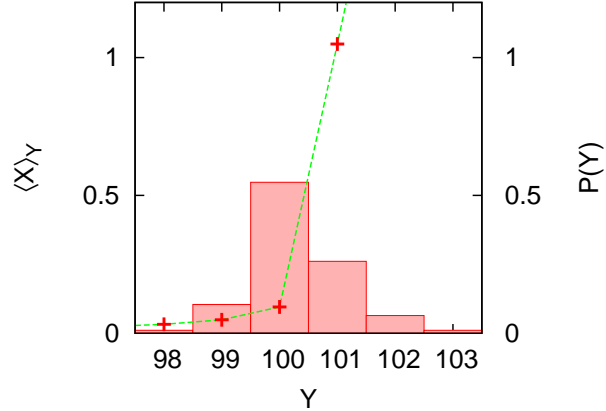}
  \caption{ The degradation rate of $Y$, $\gamma \langle X \rangle_Y$, becomes highly nonlinear for the probable region of $Y$.  The average copy number of the unbound TFs $\langle X \rangle_Y$ is shown for the different values of the total copy number ($Y$) of the TFs for the reaction process, \eqref{eqn:rate-y}.  The binding-unbinding process between the TFs and their specific promoters are assumed to be in equilibrium.   The average number of $X$ for a given value of $Y$ was estimated by using Eq.\eqref{eqn:prob-Pb}, and the probability distribution function $P(Y)$ by performing numerical simulations based on the Gillespie stochastic simulation algorithm \cite{Gillespie1977}. The above plot is for the case of $K_d = 1pM$, and $P_t = 100\mbox{nM}$ [$k_{on} = 0.167 (\simeq 1/6) (1/\mbox{nM/min})$ \cite{Elf2007}, $k_{off} = 0.000167 (1/\mbox{min})$].  } \label{fig:nonlinear-deg}
\end{figure}

Now we return to the stochastic framework and investigate how to define the retroactivity in this framework.   The stochastic fluctuations in $Y$ can be centered around the nonlinear region of $Y$ (as shown in Fig.~\ref{fig:nonlinear-deg}) such that the discreteness of $Y$ needs to be carefully taken into account.  In such a case, we cannot make a clear mathematical definition of retroactivity;  since the derivative of the degradation rate function with respect to $Y$ is not well defined.  Instead, we define the retroactivity ($R_s$) from the autocorrelation function of $Y$:
\begin{equation}
R_s  \equiv  \frac{\gamma - \gamma_a}{\gamma},
\label{eqn:Rs}
\end{equation}
where 
\begin{equation}
\gamma_a \equiv- \frac{d \log G_Y(\tau)}{d\tau},~\mbox{for the connected case,  and}
\end{equation}
\begin{equation}
\gamma \equiv- \frac{d \log G_X(\tau)}{d\tau}, ~\mbox{for the disconnected case.}
\label{eqn:gamma}
\end{equation}
The relative change of the slope of a graph of $\log G(\tau)$ vs. $\tau$ corresponds to the retroactivity.  Since the value of such defined retroactivity can deviate significantly from that of the d-retroactivity, we call this \emph{stochastic retroactivity}.

\begin{figure}[h!]
  \centering
  \includegraphics[width=3.3in]{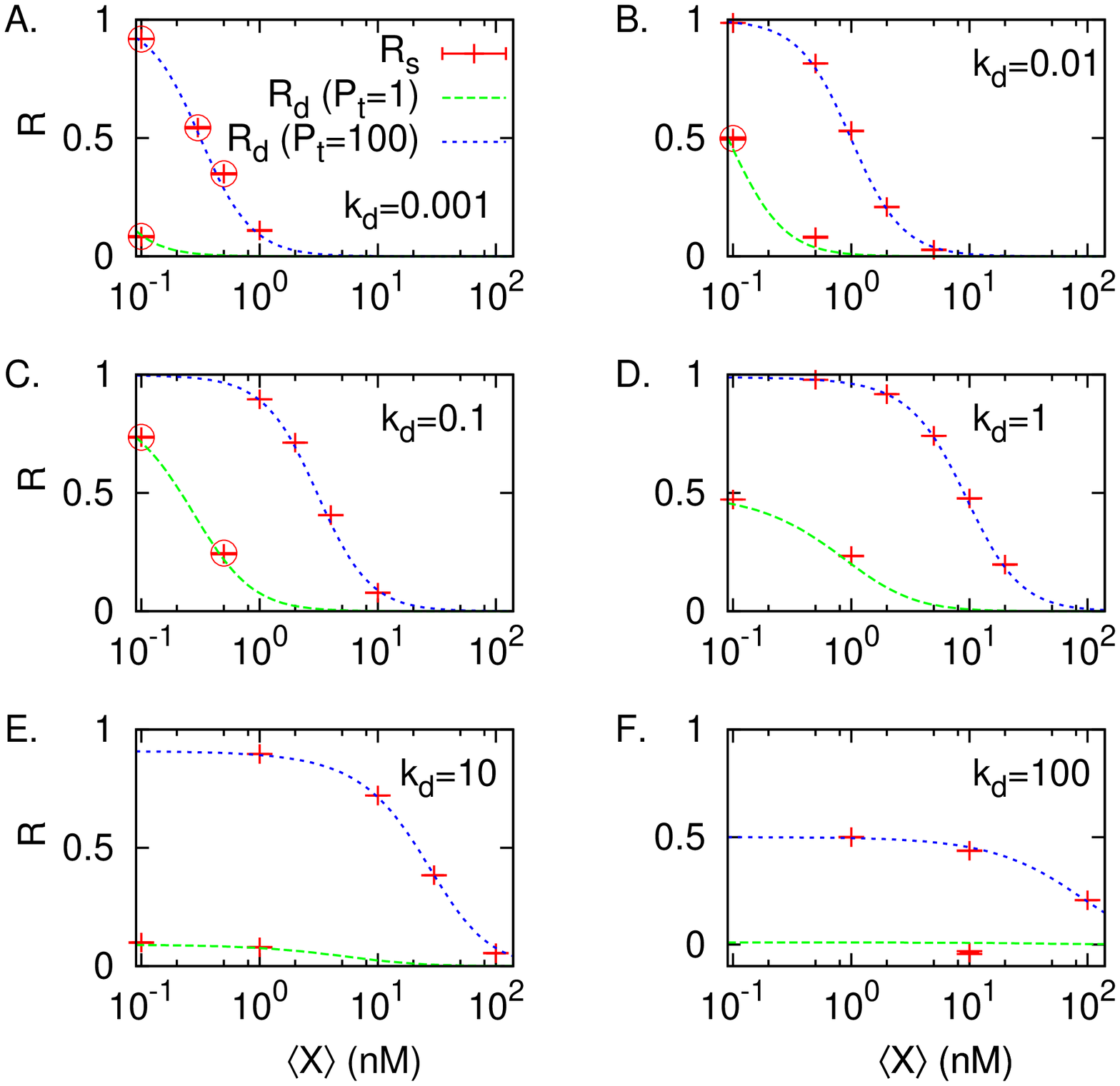}
  \caption{Retroactivity for the deterministic and stochastic cases for the different values of $K_d$ and the number of free TFs, $\langle X \rangle$. The unit of $K_d$ is nM.  The lines in the above plots correspond to d-retroactivities (Eq.~\eqref{eqn:Rd}) and the points to s-retroactivities.  Circles indicates the cases that the autocorrelation functions of $Y$ cannot be approximated to pure exponential functions as shown in Fig.~\ref{fig:acf-multi-exp} and the s-retroactivities for these cases were estimated by choosing fitting regions of $\tau$ between $0$, and $25\sim 50$ min. For the rest of the cases, the autocorrelations can be approximated to pure exponential functions for the wider fitting regions of $\tau$ between 0 and $100 \sim 250$ min.} \label{fig:R}
\end{figure}

\begin{figure}[h!]
  \centering
  \includegraphics[width=2.5in]{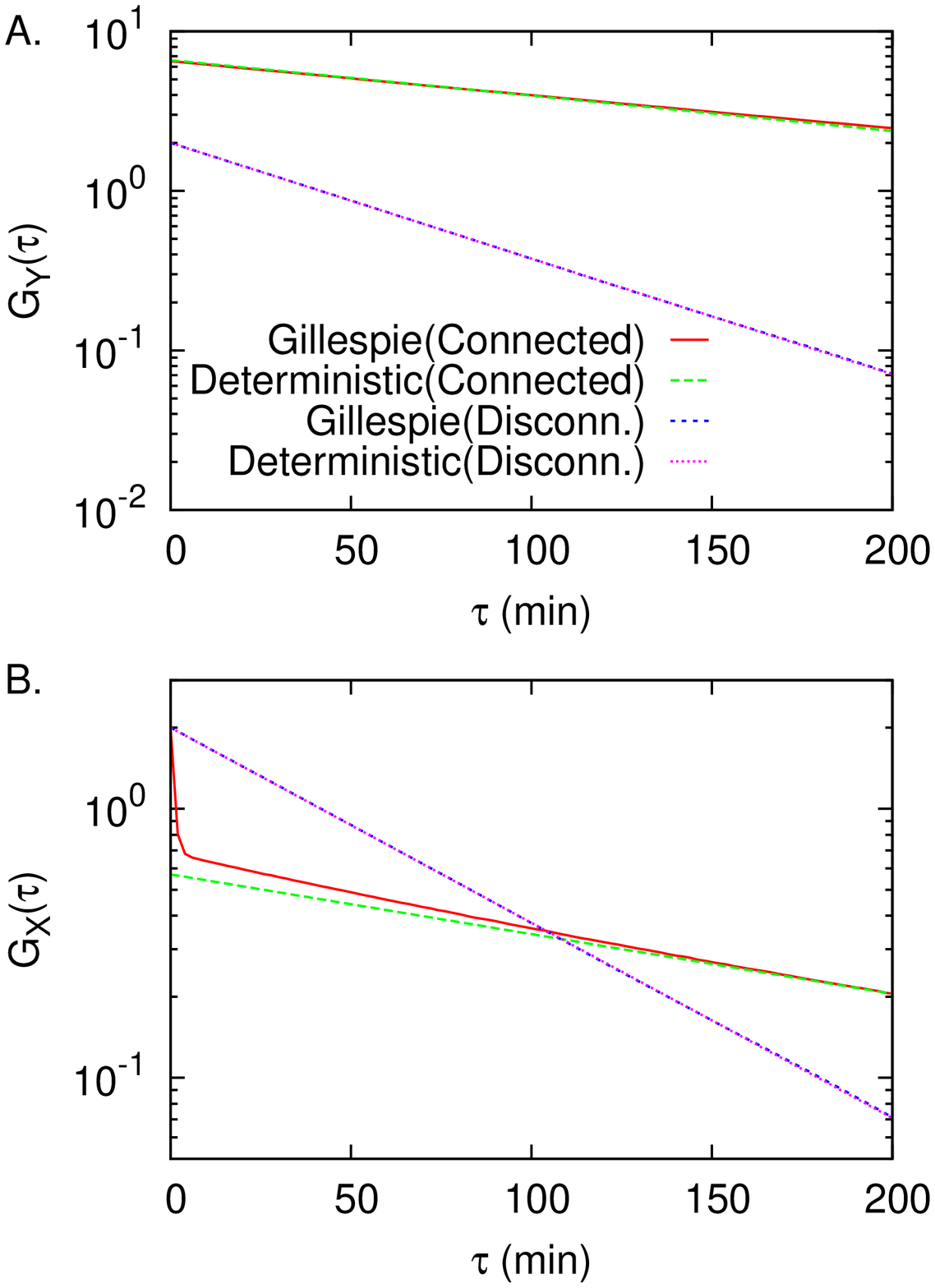}
  \caption{Autocorrelation functions of $Y$ and $X$, when a downstream module is either connected or disconnected:  The slopes of the semi-log plot of the autocorrelation functions correspond to $\gamma_a= \gamma(1-R_s)$ (Eq.~\eqref{eqn:gamma}) up to a negative sign.  The lines labeled `Gillespie' corresponds to stochastic simulation results based on the Gillespie stochastic simulation algorithm \cite{Gillespie1977}. The lines labeled `Deterministic' are exponential functions: $C \exp  \big[-\gamma(1-R_d)\tau \big]$ with $R_d$ the corresponding deterministic retroactivities Eq.~\eqref{eqn:Rd} and $C$ a normalization constant, chosen for the overlap with the `Gillespie'.  The autocorrelation functions of $X$ is not exponential (B).  The sudden drop for the small value of $\tau \sim [0,5]$ is originated from fast binding-unbinding processes.  This implies that the slow mode $Y$ of dynamics gives dominant contributions to the autocorrelation for the larger value of $\tau$.  The following  parameter values are used: $K_d = 0.1\mbox{nM}$, $\langle X \rangle = \alpha/\gamma = 2 \mbox{nM}$, and $P_t=100 \mbox{nM}$ [$\alpha = 0.0334 (\simeq 1/30)$~nM min$^{-1}$, $\gamma = 0.0167$ $(\simeq 1/60)$~min$^{-1}$, $k_{on} = 0.167$ $(\simeq 1/6)$ nM$^{-1}$min$^{-1}$  \cite{Elf2007},  and $k_{off} = 0.0167 (\simeq 1/60)$~min$^{-1}$].  } \label{fig:acf-retro}
\end{figure}

We have estimated the stochastic retroactivity by measuring the slope change in the semi-log plot of the autocorrelation functions of $Y$ for the two different processes \eqref{eqn:rate} and \eqref{eqn:rate-y}.   We have used parameter values appropriate for degradation tagged TFs in  \emph{E.~coli} host cells: the average copy number of the TF is set equal to 2 and the dissociation constant of the TF specific promoters between $0.001$ and $100$~nM and the average copy number of plasmids containing the specific promoters to $1$ and $100$ (for the reaction parameter values, refer to the caption of Fig.~\ref{fig:acf-retro} and \ref{fig:R}).  We have set the volume of  \emph{E.~coli} roughly equal to $1\mu\mbox{m}^3$, and for this volume the copy number 1 corresponds to 1~nM.  Hereafter we will interchange the unit of nM  with that of a copy number.  We have fitted the autocorrelation to an exponential function.   The measured s-retroactivity is shown to be well matched with the d-retroactivity ($R_d$) as shown in Fig.~\ref{fig:acf-retro} and \ref{fig:R}.  This result implies that the d-retroactivity can be used to estimate retroactivity for the case when stochastic fluctuations are significant, e.g., the case of low copy number TFs.  Our result shows the possibility that the insulation devices, proposed in the deterministic framework \cite{Vecchio2008} for suppressing the  retroactivity, can be applied to gene circuits, which are highly stochastic.

Dissociation constants for transcription factors and their own operators in promoter regions range from pM to several hundred nM for bacterial transcription factors.  For example, TetR proteins bind to the operator \emph{tetO} with a dissociation constant $K_d=0.001-15$nM (\emph{in vitro}) \cite{Scholz2000}, lacI with $K_d=0.1-10$pM (\emph{in vivo}) \cite{Setty2003}, luxR with $K_d=24$nM (\emph{in vitro}) \cite{Pompeani2008},  cI with $K_d=120$nM (\emph{in vivo}) \cite{Rosenfeld2005}.  For this range of the value of the dissociation constant, the stochastic retroactivity becomes non-negligible as shown in Fig.~\ref{fig:R}, implying that the autocorrelation in the output signal noise changes significantly due to the output wiring: More specifically, the time correlation in the signal noise can persist much longer after the output is wired.

In this estimation procedure for measuring the slopes, one needs to choose fitting regions of $\tau$.  Based on stochastic simulation results, the autocorrelation functions approximately become pure exponential functions for the region of small $\tau$, between 0 and $100\sim250$ min, and when the copy number of unbound TFs is larger than 1 (see Figs.~\ref{fig:acf-retro} and \ref{fig:R}).  Thus, we have fitted the autocorrelation to an exponential function for this range of $\tau$  and we will justify the reason for choosing the small value region of $\tau$ later in this Result section.

\subsection*{Retroactivity: Relative decrease in frequency bandwidth of power spectral density}

In this subsection, we will make a connection between retroactivity and the frequency response of a gene circuit.  Retroactivity will be shown to be related to the change in the frequency bandwidth of a circuit due to the circuit output wiring.  Frequency bandwidth of a circuit is defined in the deterministic framework: For example, in the process \eqref{eqn:rate},  the translation rate $\alpha$ is considered as a time-varying input signal (e.g., a sinusoidal function), and one observes the output signal of the circuit to estimate the attenuation (response) of the signal for different magnitude of frequencies.   Here in the gene circuit, the stochastic fluctuations in gene expression levels are significant, and we question ``Can one exploit the stochastic fluctuations  to understand the frequency response of the circuit?"  Simpson, et al.~performed the frequency domain analysis of noise in a negatively-autoregulated gene circuit and showed that the noise frequency spectrum (power spectral density) is significantly dependent on the negative feedback loop structure \cite{Simpson2003, Austin2006}.   Mathematically, they showed the dependence in terms of the transfer function of the feedback loop.   This means that the transfer functions, derived in the deterministic framework, can be, at least partly, identified by investigating the noise frequency spectrum.     Therefore, we will investigate the noise frequency spectrum to investigate the change in frequency response due to the circuit wiring.  

\begin{figure}[h!]
  \centering
  \includegraphics[width=3.3in]{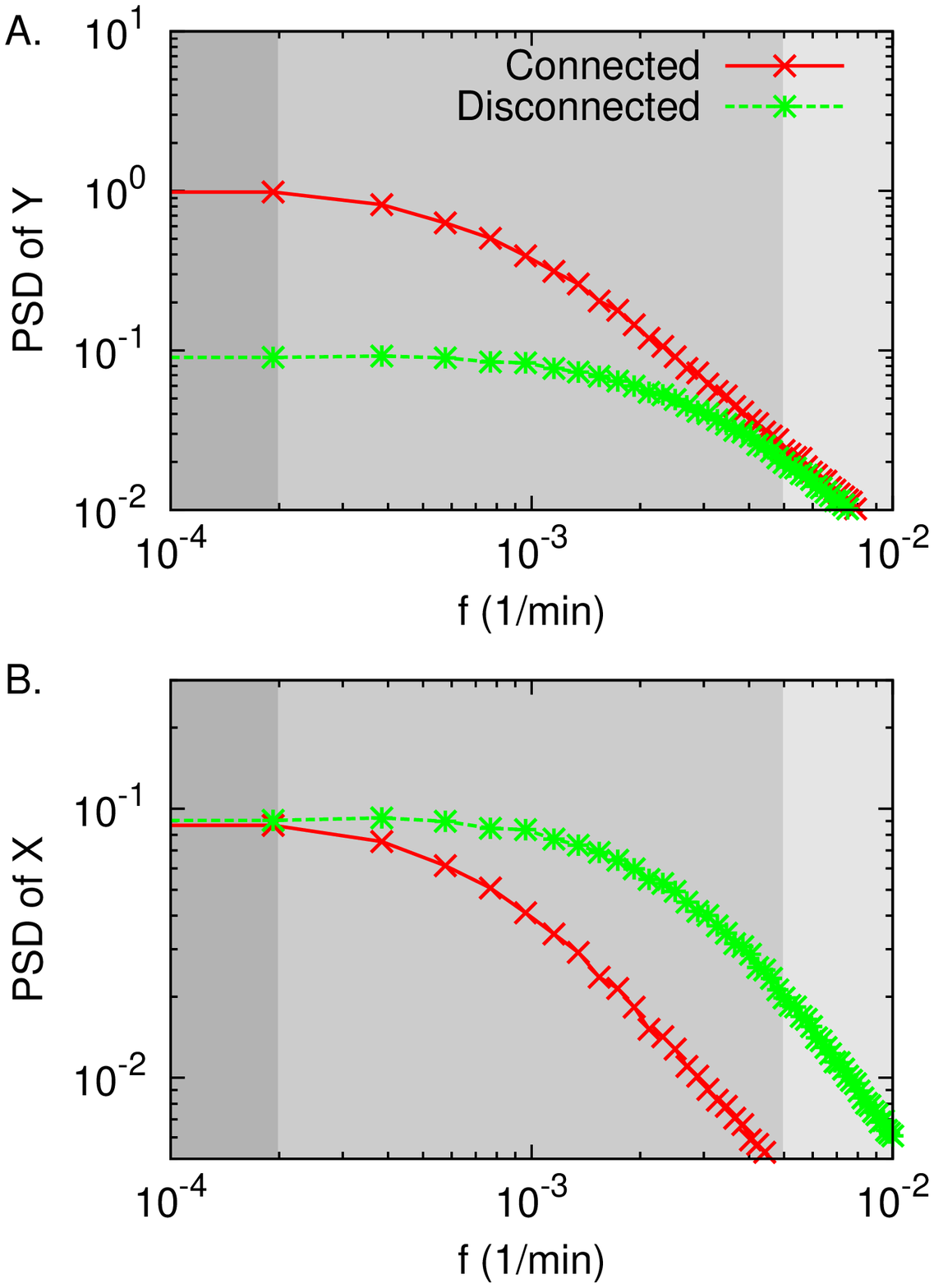}
  \caption{Power spectral densities (PSD) (one-sided) obtained from the autocorrelation functions of $Y$ and $X$, respectively shown in Fig.~\ref{fig:acf-retro}A and B.  The low frequency components of signal $X$ are not affected by the connection to a downstream component (B), while the high frequency components of $Y$ are not (A). The threshold point, where the PSD starts to drop down, is shifted to the left for both $X$ and $Y$, due to the connection.  Three regions of frequency colored above correspond, respectively  from left to right,  to low frequency region where the signal $X$ is not affected while $Y$ is, intermediate region where both $X$ and $Y$ are affected, and high frequency region where $Y$ is not affected while $X$ is.  } \label{fig:psd}
\end{figure}

We perform the Fourier transform of the autocorrelation function of $Y$, resulting in a (two-sided) power spectral density.  If the estimated autocorrelation function of $Y$ can be well approximated to a pure exponential given by $W_I (2 \gamma_a)^{-1} \exp(-\gamma_a |\tau|)$ (in the previous section, we assume $\tau \geq 0$, but here $\tau$ can be negative), its Fourier transform is given as 
\begin{equation}
\tilde{G}_Y(\omega) \simeq \frac{1}{2 \pi}\frac{W_I}{\gamma_a^2+\omega^2}.
\label{eqn:psd}
\end{equation}
This power spectral density (PSD) decreases by half from its maximum when $\omega$ is equal to the apparent degradation rate constant, $ \gamma_a$, and thus this value is called the bandwidth of the power spectrum of $Y$.  This indicates that the bandwidth  decreases from $\gamma$ to $\gamma_a = \gamma(1-R_s)$ due to retroactivity when two modules are connected (see Fig.~\ref{fig:psd}A and B).   The relative decrease in the bandwidth can also be used to define the retroactivity.  

Finally, we discuss how to reduce the change in the frequency bandwidth, i.e., retroactivity.    Del Vecchio et al. proposed two mechanisms for this: a negative feedback on $X$ and the amplification in the degradation rate of $X$ \cite{Vecchio2008}.     More specifically, the negative feedback increases the apparent degradation rate constant since any perturbation in $X$ will damp away more strongly due to the feedback, and the amplification of the degradation rate directly increase the degradation rate constant.   Thus, these mechanisms can reduce the retroactivity.

\subsection*{Modularity in circuit dynamics and two choices of output signals: X or Y?}
Figure \ref{fig:psd} shows very interesting aspects of the frequency responses in the circuit.  The noise spectrum of signal $Y$ (the total number of the transcription factor) changes significantly at low enough frequency while it does not at high enough frequency, but that of signal $X$ (the number of the unbound transcription factor) shows the opposite behavior.  This implies that the signal $X$ becomes not affected by wiring when the circuit is operated at the low frequency, and the signal $Y$ does not when operated at the high frequency, although the retroactivity is still causing the significant change in the frequency bandwidth.
This means that circuit dynamics can be investigated in a modular way by choosing an appropriate output signal, either $X$ or $Y$, depending on the magnitude of the operating frequency.  

\begin{figure}[h!]
  \centering
  \includegraphics[width=3.in]{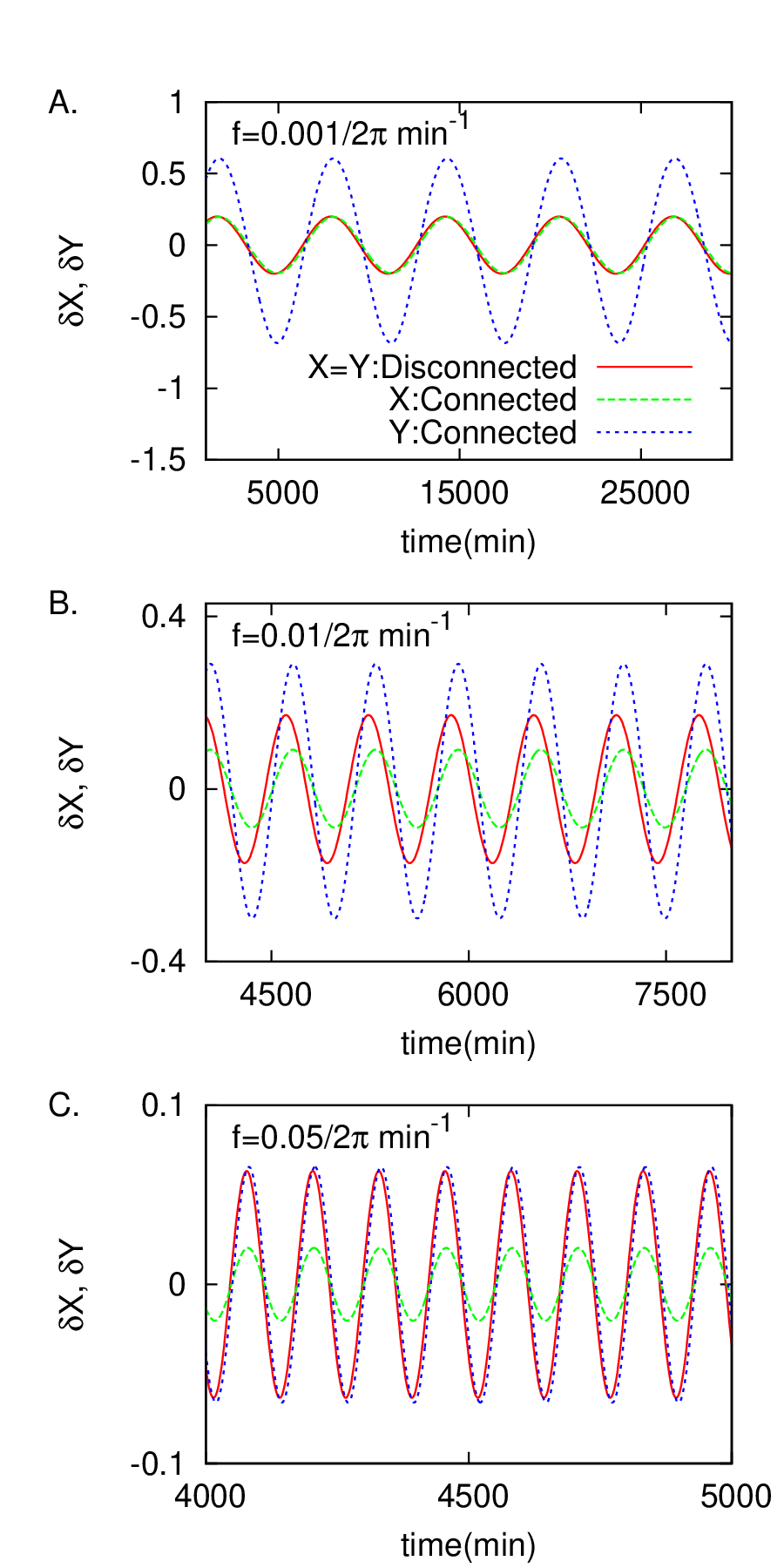}
  \caption{Signals $X$ and $Y$ are compared for different values of frequency when process \eqref{eqn:rate} is deterministically simulated.  The translation rate $\alpha$ is given as a time-varying signal with a harmonic oscillation as shown in Eq.~\eqref{eqn:alpha}.  The frequency of the oscillation is chosen from the three regions shown in Fig.~\ref{fig:psd}. The parameter values are identical to those used in  Fig.~\ref{fig:acf-retro}. }  
\label{fig:XYsignal}
\end{figure}

To validate the above prediction, we simulate the process~\eqref{eqn:rate} deterministically, with $\alpha$ allowed to oscillate at different frequencies, which are chosen from three different regions indicated in Fig.~\ref{fig:psd}:
\begin{equation}
\alpha(t) = 1+0.1 \sin ( 2 \pi f t)
\label{eqn:alpha}
\end{equation}
 with $f = 0.001/2\pi$, $0.01/2\pi$, and $0.05/2\pi$ (see Fig.~\ref{fig:XYsignal}).  Output signals $X$ and $Y$ show oscillations with basal levels: $X(t) = \bar{X}+ \delta X(t)$ or $Y(t) = \bar{Y} + \delta Y(t)$.   We compare the oscillatory components ($\delta X$ and $\delta Y$) of the output signals for the two different cases when the output is connected and disconnected  (Fig.~\ref{fig:XYsignal}).  At the low frequency $f=0.001/2 \pi$, $\delta X$ does not change due to wiring while $\delta Y$ does.  At the high frequency $f=0.05/2 \pi$, $\delta X$ changes due to wiring while $\delta Y$ does not.  At the intermediate frequency $f=0.01/2 \pi$, both the signals $X$ and $Y$ change.  This verifies the prediction based on power spectral densities,  and implies that the modularity of gene circuits can be tested by observing gene expression fluctuations.

Now, we will explain in detail the origin of this different behavior of $X$ and $Y$.  First, we investigate the case that the circuit operates in the low frequency region.  The mean level of $X$, corresponding to zero frequency component,  is determined by its synthesis and degradation rate because binding and unbinding reactions are balanced ($\langle X \rangle =  \alpha/\gamma$).  The same argument is applied for the low frequency region ($\omega$)  colored in dark gray in Fig.~\ref{fig:psd}, where the balance still holds, and the two PSD of $X$ when connected and disconnected  are almost the same as each other. This means that the low frequency components of signal $X$ is not distorted by the wiring.   Therefore, when the dynamics of the individual modules are described by $X$, the signal distortion in $X$ due to the retroactivity can be minimized by operating at the low frequency. 

We consider the case that the circuit operates in the high frequency region.  In this case, we investigate the signal $Y$ rather than $X$, because the two PSDs of $Y$ when connected and disconnected converges to each other at the high frequencies as shown in Fig.~\ref{fig:psd}.  More specifically, the PSD of $Y$ becomes $W_I/2 \pi \omega^2$ as $\omega \gg \gamma > \gamma_a$ by using Eq.~\eqref{eqn:psd}.  This implies that when the dynamics of the modules are described by $Y$, the signal distortion in $Y$ due to the retroactivity can be minimized when the circuit is operated at the high frequency.  We note that the PSD of $Y$ at the low frequencies changes due to the wiring because there is an extra amount of bound TFs when wired.  This PSD study implies that modular description is possible by choosing the right signal variables either $X$ or $Y$ depending on the operating frequencies.

\subsection*{Non-linear degradation rate of $Y$}
Non-linear degradation rate of $Y$ causes the autocorrelation function to be non-exponential.
First, we discuss why the degradation rate of $Y$ becomes non-linear. We have assumed that the degradation of bound TFs is negligible compared to that of unbound TFs.  When the copy number of the total TFs is much larger than that of the specific promoters, most of the TFs are found to be unbound, and mathematically this is expressed as $Y\simeq X$.  Thus, the degradation rate of $Y$ ($= \gamma X$) becomes approximately proportional to $Y$: $\gamma X \simeq \gamma Y$.  However, if the copy number of the total TFs is comparable to that of the specific promoters, most of the TFs become bound to the promoter regions.  Since these bound TFs degrade very slowly, the net degradation rate of the total TFs becomes significantly reduced (Fig.~\ref{fig:nonlinear-deg}). Thus, the degradation rate of $Y$ is not proportional to $Y$. 

\begin{figure}[Ht]
  \centering
  \includegraphics[width=3.3in]{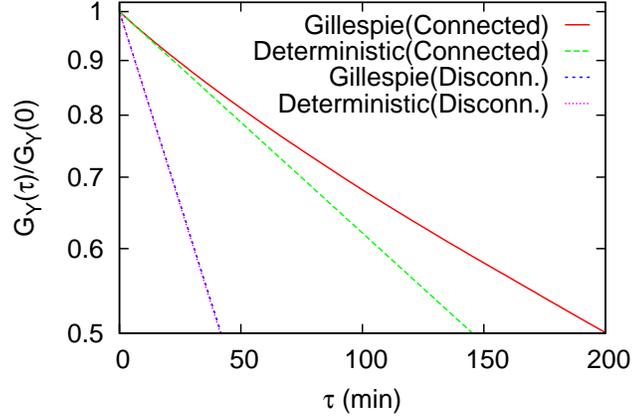}
  \caption{Semi-log plot of normalized autocorrelation functions with non-exponential shapes. Parameters: $K_d = 0.1\mbox{nM}$, $\langle X \rangle = \alpha/\gamma = 0.1 \mbox{nM}$, and $P_t=1 \mbox{nM}$ [$\alpha=0.00167 \mbox{nM/min}$, $\gamma = 0.0167 (\mbox{1/min})$, $k_{on} = 0.167 (\mbox{1/nM/min})$, $k_{off} = 0.0167 (\mbox{1/min})]$. For the explanations on the legend, refer to Fig.~\ref{fig:acf-retro}  } \label{fig:acf-multi-exp}
\end{figure}

Such nonlinearity causes the autocorrelation function of $Y$ not to be a pure exponential function as shown in Fig.~\ref{fig:acf-multi-exp} (while the linear degradation rate causes the autocorrelation function to be a pure exponential function as we have discussed previously).  The reason of the non-exponential autocorrelation is that the fluctuations in $Y$ are composed of many modes of fluctuations with different correlation time constants.  The slower modes among them can persist its correlation for the longer time interval, contributing more dominantly to an autocorrelation function for the region of large $\tau$.  Consider the case of the nonlinear degradation shown in Fig.~\ref{fig:nonlinear-deg}.  Most of the fluctuations occur between 99 and 102.  Among them, the fluctuations between 100 and 102 will have the correlation time constant $\simeq 1/\gamma$, while those between 99 and 100 will have the correlation time constant much larger than $1/\gamma$.  Thus, the slope of the autocorrelation (in the semi-log plot) becomes tilted upward for large values of  $\tau \gtrsim 200$ as shown in Fig.~\ref{fig:acf-multi-exp}.  This slope change,  which typically appears in our simulations, becomes negligible for the case that $\langle X \rangle >1$ for the time interval of less than several hours (see Fig.~\ref{fig:R}).

\subsection*{Fitting regions of $X$ for s-retroactivity measurements}
The remaining question is to be `why is the slope in the small $\tau$ region (e.g., [0,50] in Fig.~\ref{fig:acf-multi-exp}) related to the d-retroactivity?' To answer this question, we need to discuss how the deterministic retroactivity can be estimated.  In \cite{Vecchio2008}, the retroactivity was measured (in simulations) either by applying static or oscillatory perturbations in the TF creation rate.  The strength of the perturbations needs to be small enough that the  degradation rate of the total TFs changes in proportion to the perturbation strength.  To make a correspondence to the deterministic case, in the stochastic framework, we have to consider small-amplitude fluctuation components of $Y$ to estimate the retroactivity.  If a system is not multi-stable, small-amplitude fluctuations occur more frequently than large-amplitude fluctuations.  To measure the contribution of the dominant fluctuations (which are the small-amplitude ones), we need to consider the autocorrelation for the sufficiently small value of $\tau$; for the small value of $\tau$, all the fluctuation components, with relaxations either slow or fast, can contribute to the autocorrelation and thus the major contribution comes from the fluctuation components occurring most frequently, which are the small-amplitude fluctuations.  This is why we have chosen the region of the small value of $\tau$ for the estimation of the retroactivity.  By choosing the fitting region of $\tau$ sufficiently small enough (between 0 and 25$\sim$50 min), the s-retroactivity is shown to be well matched with the d-retroactivity for the reasonable parameter values for experiments.

\subsection*{Computational approach for estimating modularity from gene expression fluctuations}
In this subsection, we show how to computationally estimate the level of modularity based on fluorescence data from single cell experiments.  Synthetic circuits are often transferred to host cells, e.g., \emph{E.~coli}, which duplicate themselves by completing cell cycles.  This causes the intra-cellular concentrations to fluctuate.  In addition, the host cells are under other unidentified extrinsic noise sources.  Such extrinsic noise has been shown to affect the autocorrelation functions \cite{Rosenfeld2005, Austin2006, Weinberger2008, Dunlop2008} and needs to be taken into account for estimating the level of modularity. 

If a transcription factor with a  fluorescence marker is tagged for degradation, the lifetime of the TF can be comparable to the cell doubling time ($T_d$).   Then, the autocorrelation function of the fluorescence emitted from the TF can be fitted to the following function: 
\begin{equation}
G_Y(\tau) = A e^{-\delta_E \tau} + B e^{-(\delta_E + \gamma_a)\tau},
\label{eqn:autocorr1}
\end{equation}
where $\delta_E=\log(2)/T_d$ with $T_d$ the cell doubling time and $\gamma_a$ the apparent degradation rate constant, defined in Eq.~\eqref{eqn:gamma}.  The above form of the autocorrelation has been investigated in its Fourier transform (power spectral density)  by Austin \emph{et al.} by using a plasmid containing a GFP variant with its half-life reduced \cite{Austin2006}.  They did not consider the retroactivity dependence in $\gamma_a$.  
The power spectral density, a Fourier transform of the autocorrelation function Eq.~\eqref{eqn:autocorr1},  is obtained as \cite{Austin2006}:
\[
\tilde{G}(\omega) = \frac{2A \delta_E}{\delta_E^2+\omega^2}+\frac{2B (\delta_E+\gamma_a)}{(\delta_E+\gamma_a)^2+\omega^2}.
\]
Depending on the relative magnitude of $A$ and $B$, the bandwidth is determined by either $\delta_E$ and $\delta_E + \gamma_a$ or both.  

We propose the following data analysis procedures, to estimate retroactivity:
\begin{itemize}
\item[1.] Obtain single cell fluorescence trajectories by connecting each trajectory belonging to a cell lineage, by following the method described in \cite{Austin2006}.
\item[2.] Estimate the autocorrelation function of the fluorescence.
\item[3.] Perform a non-linear fit by using Eq.~\eqref{eqn:autocorr1} to estimate $\delta_E$ and $\gamma_a$.  Alternatively, estimate $\delta_E = \log(2)/T_d$ by measuring cell doubling time ($T_d$) from the cell lineage, and then perform a non-linear fit to estimate $\gamma_a$ by using Eq.~\eqref{eqn:autocorr1}.  
\item[5.] Estimate the retroactivity: Repeat the above procedures in the connected and disconnected case and obtain the relative change in the apparent degradation rate constants  by using Eq.~\eqref{eqn:Rs}.
\end{itemize}

We can identify the frequency regions that the circuits operate in a modular fashion:
\begin{itemize}
\item[1.] Obtain power spectral densities by applying the Fourier transformations on the estimated autocorrelation functions.  
\item[2.] Identify the region of frequency where the two PSDs obtained in both the connected and disconnected cases overlap each other.  This is the region of frequency where signal $Y$ is preferably chosen for the modular analysis.   
\item[3.] Identify the region of frequency where the two PSDs of $Y$ becomes parallel to each other.  This is the region of frequency where signal $X$ is chosen.
\end{itemize}

We present a model for gene expression in the Appendix where we derive  the above autocorrelation function Eq.~\eqref{eqn:autocorr1}, which will be shown to be consistent with the current experimental measurements \cite{Rosenfeld2005, Austin2006, Dunlop2008, Weinberger2008}.   The model process is described in the Langevin dynamics framework, with the retroactivity effect taken into account.  In the derivation of Eq.~\eqref{eqn:autocorr1}, we neglect the reaction processes of fast relaxation ($\lesssim 10$ min) due to transcription, and mRNA-degradation \cite{Austin2006}.

If the copy number of unbound TFs is very low $\langle X \rangle \lesssim 1$, the autocorrelation may not be a pure exponential due to both the extrinsic noise and the nonlinearity of the degradation rate of $Y$.  In this case, the fitting region becomes comparable to the fast relaxation time scale and we need to take into account the contribution of the fast time scale dynamics to the autocorrelation functions (shown in Eq.~\eqref{eqn:autocorr}). The fast time scale dynamics is, however, not well understood compared to the slow time scale dynamics \cite{Rosenfeld2005, Austin2006, Dunlop2008, Weinberger2008}.  In addition, taking into account the fast time scale dynamics increases the number of fitting parameters.  Therefore, an accurate estimation can be less promising in the case for the low copy number of the unbound TF.

\subsection*{If the degradation process of bound TFs is not negligible, retroactivity becomes reduced.}
In this subsection, we take into account  the degradation of bound transcription factors.  The total copy number of the transcription factors ($Y$) is still a pure slow mode because the degradation of the bound transcription factors may not be faster than the unbound ones.  The degradation of the bound TFs ($P_b$) contributes to the degradation of $Y$: 
\begin{equation*}
\xrightarrow{~~\alpha~~} Y \xrightarrow{\gamma X + \gamma_b P_b}, ~~~\mbox{i.e.}~~~ \xrightarrow{~~\alpha~~} Y \xrightarrow{\gamma \big[ Y - (1-\gamma_b/\gamma) P_b \big]} 
\end{equation*}
with $\gamma_b$ the degradation rate constant of the bound TFs and for the  reaction process shown on the right,  we used $X=Y-P_b$.  
When the bound TFs degrades at the same speed as the unbound TFs, the degradation rate becomes proportional to $Y$ with its rate constant equal to $\gamma$, resulting in zero retroactivity.  For the case of $\gamma \neq \gamma_b$, however, the apparent degradation rate constant becomes
\[
\gamma_a = \gamma (1-(1-\gamma_b/\gamma)R),
\]
where $R$ is the retroactivity in the case that the degradation of the bound TFs is negligible (Eq.~\eqref{eqn:gamma} or \eqref{eqn:Rd}).  Thus, the retroactivity is expected to decrease by a factor of $1-\gamma_b/\gamma$ when the degradation of the bound TFs are allowed.

\subsection*{Summary}
In this paper, we have considered a gene transcription-translation module and have investigated how noise correlations in the circuit signals change after the output of the circuit module is wired to another module.  We call such changes in noise correlation, stochastic retroactivity.  More specifically, the retroactivity was defined from the autocorrelation functions of gene expression levels of transcription factors (TFs).  When the TFs can bind specifically to downstream promoters, the autocorrelation is shown to decrease more slowly compared with the case when not wired.  This implies that the noise correlation persists longer and that the module filters out signal components corresponding to high frequencies more strongly.  

We have made a connection between the stochastic retroactivity and the retroactivity investigated in the deterministic framework \cite{Vecchio2008}.  For reasonable parameter values for experiments, our simulation study shows that the autocorrelation functions approximately become pure exponential functions (if we neglect extrinsic noise) and the estimated stochastic retroactivity is well matched with the deterministic retroactivity.  

We have shown that the retroactivity is related to the amount of decrease of the frequency bandwidth of the power spectral density of the gene expression.  We have proposed that depending on the magnitude of an  operating frequency, the circuit dynamics can be analyzed in a modular fashion by choosing the appropriate circuit output signal: at low enough frequencies, the concentration of the free TF, and for high enough frequencies,  the concentrations of the total TF.

This autocorrelation study provides a experimental method for measuring the retroactivity to test the level of modularity in gene circuit dynamics.  This will eventually help for the modular design of the gene circuits with elaborate dynamical functionality.

\section*{Materials and Methods}
\subsection*{Gene expression models and autocorrelation functions}
In this section, we present a stochastic model of gene expression for transcription factors (TFs), taking into account retroactivity due to binding-unbinding interactions between the TFs and their specific promoter region.  We aim to describe the dynamics of the processes in the slow time scale, where we assume the binding-unbinding processes of the TFs to be in equilibrium.  In the slow time scale, the dynamics can be described by a slow variable, the total number of TFs ($Y$), including bound and unbound TFs.   We assume that the translation of the TF occurs at a constant rate and the degradation at a rate proportional to $Y$ with a apparent degradation rate constant, $\gamma_a$. The reason for the latter assumption is based on the simulation result that its autocorrelation can be approximated to be a pure exponential function for the case that the average number of the unbound TFs is larger than one.  Then, we can model the process by the following Langevin equation: 
\begin{equation}
\frac{d Y}{d t} = \alpha - \gamma' Y + I + E,
\label{eqn:ode}
\end{equation}
where $\gamma'$ is the sum of the apparent degradation rate constant ($\gamma_a$) and dilution rate constant ($\delta_E$):
\[
\gamma' \equiv \gamma_a+ \delta_E.
\]  
I and E represent intrinsic and extrinsic noise, respectively, which are exponentially-correlated  \cite{Austin2006, Dunlop2008, Weinberger2008}:
\[
\langle I(t) I(t+\tau) \rangle = W_I\exp(-\tau/T_I),
\]
and
\[
\langle E(t) E(t+\tau) \rangle = W_E\exp(-\tau/T_E),
\]
where the angle bracket denotes an average over a time series at the stationary state.  

To obtain the  autocorrelation function of $Y$,  we obtain the integral-type solution of Eq.~\eqref{eqn:ode} for $Y(t)$ and substitute the solution into Eq.~\eqref{eqn:autocorr-def}, resulting in:
\begin{eqnarray}
\lefteqn{G_Y(\tau) = \frac{W_I \delta_I}{\delta_I^2-\gamma'^2}\Big( - \frac{e^{-\delta_I \tau}}{\delta_I} + \frac{e^{-\gamma' \tau}}{\gamma'} \Big) } \nonumber \\
&&~~~~~+ \frac{W_E \delta_E}{\delta_E^2-\gamma'^2}\Big( - \frac{e^{-\delta_E \tau}}{\delta_E} + \frac{e^{-\gamma' \tau}}{\gamma'} \Big).
\label{eqn:autocorr}
\end{eqnarray}

We will show that the above autocorrelation function, Eq.~\eqref{eqn:autocorr}, is consistent with the functional forms obtained or used in \cite{Rosenfeld2005, Austin2006,  Dunlop2008}.  To show the consistency with Austin et al. \cite{Austin2006}, we consider that the internal noise correlation time is much smaller than the extrinsic noise correlation time (i.e., $\delta_I \gg \delta_E$).  We can simplify the above equation to Eq.~\eqref{eqn:autocorr1}, giving an identical result presented in \cite{Austin2006} up to the normalization constant.   In \cite{Rosenfeld2005, Dunlop2008}, the authors did not use degradation-tagged TFs in measuring the expression level.  The life time of the TFs was much larger than the extrinsic noise correlation time ($\delta_E\gg \gamma \gtrsim \gamma_a$, i.e., $\gamma' \simeq \delta_E$).  In this case, the autocorrelation can be simplified to $A' \exp(-\delta_I \tau) + B' \exp(-\delta_E \tau)$ (where internal noise correlation is not ignored) \cite{Rosenfeld2005, Dunlop2008}.  Therefore, this consistency supports the Langevin dynamics description Eq.~\eqref{eqn:ode} for gene expression, taken into account the retroactivity,  are valid within the accuracy of current experimental data.

\subsection*{Slow mode ($Y$) dynamics}
In this section, we justify the reaction process of $Y$, described by Eq.~\eqref{eqn:rate-y1}, and derive the average copy number of the bound transcription factor, $\langle P_b \rangle_Y$.  The TF degradation is assumed to be much slower than the equilibration time of the binding-unbinding process between the TF and its specific promoter regions: $k_{on}X + k_{off} \gg \gamma$, cf. \cite{Kepler2001, Simpson2004, Vecchio2008}.  

We construct the master equation for $Y$ and $P_b$:
\begin{eqnarray*}
\lefteqn{\frac{\partial P(Y,P_b)}{\partial t} = P(Y-1,P_b) \alpha}\\
&& + P(Y, P_b-1) k_{on}(P_{t}-P_b+1) (Y-P_b+1)\\
	&&+P(Y+1,P_b) \gamma (Y+1-P_b) + P(Y,P_b+1) k_{off}(P_b+1)\\
	&&-P(Y,P_b) \big[ \alpha + \gamma X + k_{off} P_b + k_{on}(P_{t}-P_b)X \big].
\end{eqnarray*}
As in the deterministic case, we assume the transcription factor binding-unbinding process is in a quasi-equilibrium state, conditioned on the slow variable, i.e., $P(Y,P_b;t)  = P(P_b|Y) P(Y;t)$ \cite{Rao2002}. This approximation simplifies the above equation to another master equation for the slow variable $Y$:
\begin{eqnarray*}
\lefteqn{\frac{\partial P(Y;t)}{\partial t} = \alpha\big[P(Y-1;t)-P(Y;t)\big] }\\
	&&+ \gamma \big(Y+1-\langle P_b \rangle_Y \big)P(Y+1;t) - \big( Y-\langle P_b \rangle_Y \big)P(Y;t).
\end{eqnarray*}
This means that the slow process can be simplified to the reaction process described by Eq.~\eqref{eqn:rate-y1}.

Under the equilibrium approximation, the fast mode process satisfies the detailed balance:
\[
k_{off}(P_b+1) ~P(P_b+1|Y) = k_{on}(P_{t}-P_b)(Y-P_b)~P(P_b|Y).
\]
This enables one to obtain the equilibrium probability distribution function of $P_b$. We obtain the average copy number of the bound transcription factor, $\langle P_b \rangle_Y$:
\begin{eqnarray}
\langle P_b \rangle_Y &=& \sum_{P_b=0}^Y  \Big(\frac{k_{on}}{k_{off}}\Big)^{P_b} \frac{P_{t}! Y!}{(P_b-1)! (P_{t}-P_b)! (Y-P_b)!}\nonumber \\
&&\times ~P(0|Y), 
\label{eqn:prob-Pb}
\end{eqnarray}
where $P_t$ denotes the total number of the promoters, $P_f+P_b$.

\section*{Acknowledgments}
This work was supported by a National Science Foundation (NSF) Grant in Theoretical Biology 0827592. Preliminary studies were supported by funds from NSF FIBR 0527023.  The authors acknowledge useful discussions with Hong Qian.

\bibliography{../../research}

\begin{thebibliography}{10}
\bibitem{note-retroact}
The determinic evolution of the concentration of $X$ is governed by $dX/dt =
  \alpha - \gamma X$. Its solution is given by $X(t)=\left( X(0) -
  \alpha/\gamma \right) e^{-\gamma t} + \alpha/\gamma$, with $X(0)$ the initial
  concentration. Thus, the exponential function decreases sufficiently at $t
  \simeq 1/\gamma$.
\bibAnnoteFile{note-retroact}
\end{thebibliography}

\pagebreak


\end{document}